\documentclass[aps,amsfonts,amsmath,prd,preprint,nofootinbib]{revtex4}
\newcommand{\beq}{\begin{equation}}
\newcommand{\eeq}{\end{equation}}

\usepackage{epsfig,bbm,cancel,ulem}
\usepackage{xcolor}
\usepackage[breaklinks=true]{hyperref}
\usepackage{graphicx}
\usepackage{amsmath}
\usepackage{amssymb}
\DeclareMathOperator{\Tr}{Tr}

\begin{document}

\title{First-order formalism for Alice string}

\author{E.~Acalapati}
\email{muhammad.ezra81@ui.ac.id}
\author{H.~S.~Ramadhan\footnote{Corresponding author.}}
\email{hramad@sci.ui.ac.id}
\affiliation{Departemen Fisika, FMIPA, Universitas Indonesia, Depok, 16424, Indonesia. }
\def\changenote#1{\footnote{\bf #1}}

\begin{abstract}
We apply the {\it first-order formalism} method to obtaining BPS equations for Alice string. This is done by generalizing the well-known first-order formalism to the case of non-Abelian strings. We do not assume any specific gauge group nor the shape of the kinetic term function, but require only that the fields are axially-symmetric and static.  With this formalism we reproduce the BPS equations of $SU(2)\times U(1)$ Alice strings~\cite{Chatterjee:2017jsi}, and present their corresponding numerical solutions.
\end{abstract}

\maketitle
\thispagestyle{empty}
\setcounter{page}{1}

\section{Introduction}

In field theory, the Bogomolny-Prasad-Sommerfield (BPS)\footnote{Also equivalently called the Bogomolny equations.} equations are a very convenient way of solving field equations for certain configurations of topological solitons, since they reduce the second-order Euler-Lagrange equations into first-order~\cite{Bogomolny:1975de, Prasad:1975kr}. Physically, the the solutions represent the lowest global energy configuration and thus stable. The original Bogomolny mechanism~\cite{Bogomolny:1975de} relies so much on ``trial-and-error" method. Although successful for the standard canonical defects, it is of little help in the face of various non-canonical or exotic non-standard defects. To overcome this difficulty, several formalism have been proposed to build systematic algorithm for obtaining the BPS equations. Bazeia, Losano, and Menezes observed that stability in static field theory implies the {\it stressless condition}, $T_{ij}=0$~\cite{Bazeia:2008tj, Bazeia:2007df}. In~\cite{Adam:2013hza} the authors generalize the notion of self-duality to construct the BPS equations and their corresponding topological charge. One of us (HSR) and collaborators showed that the BPS condition can be derived directly from the Euler-Lagrange equations via separation of variable without the need to appeal to the Hamiltonian; we dubbed the {\it on-shell method}~\cite{Atmaja:2014fha, Atmaja:2015lia}. In~\cite{Atmaja:2015umo}, Atmaja suggests a method for obtaining BPS equations by relying on the so-called {\it BPS energy function} $Q$. This method is suitable for BPS defects with non-canonical kinetic terms. In~\cite{Bazeia:2017nas} the authors developed the so-called {\bf first-order formalism}, which was based on the concept of strong necessary condition~\cite{Sokalski:2001wk}, to obtain BPS equations for generalized vortices. Interestingly, they did not start from the stressless condition as in~\cite{Bazeia:2008tj, Bazeia:2007df}. Rather, by assuming axial symmetry of the energy-momentum tensor, they show that the stressless condition is the result of stability under energy rescaling.

The first-order formalism~\cite{Bazeia:2017nas} was constructed for Abelian vortices. It is not surprising since topologically stable vortices exist when the first homotopy group of the corresponding vacuum manifold is nontrivial, $\pi_1\left(\mathcal{M}\right)\neq0$. The simplest group that satisfies this condition is when a $U(1)$ symmetry spontaneously broken by a complex scalar field, thus Abelian. On the other hand, vortices (or strings) can also be shown to exist in certain non-Abelian theories. Perhaps the simplest non-Abelian string is the $Z_N$-strings, discussed by Nielsen and Olesen in the very same paper that pointed out the existence of Abelian vortex~\cite{Nielsen:1973cs}. This string can be formed through a maximal breaking of $SU(N)\rightarrow Z_N$. Within the the context of Super QCD, the BPS equations for $Z_N$ strings were obtained in~\cite{Kneipp:2001tp, Kneipp:2002yv, Kneipp:2002gk, Kneipp:2008dc}. Note that non-Abelian strings can exist despite topologically trivial. Vachaspati and Achuccaro show that string coming from the breaking of $SU(2)_g\times U(1)_l\rightarrow U(1)_g$ can be stable even though the vacuum manifold is $S^3$~\cite{Vachaspati:1991dz}, dubbed the {\it semilocal} string. This is due fact that there is gradient energy cost to consider when deforming it to vacuum. When the $SU(2)$ is gauged the defects become {\it electroweak} string~\cite{Vachaspati:1992jk}. The readers are advised to consult Vilenkin and Shellard~\cite{Vilenkin:2000jqa} for a more complete discussion on non-Abelian strings.

Four decades ago, Schwarz found a class of non-Abelian vortices, dubbed the {\it Alice strings}, where the unbroken symmetries become multi-valued when parallel-transported around it~\cite{Schwarz:1982ec, Benson:2004ue}. This happens because the unbroken group $H$ contains both the usual $U(1)$ electrodynamics (with charge operator $Q$) and an element $X$ of the disconnected component of $H$ such that 
$XQX^{-1}=-Q$~\cite{Bucher:1991qhl}. This property leads to a peculiar behavior: a test particle shall flip its charge's sign after circumnavigating the string once. Another exotic feature of this defects is that when the string forms a closed loop it can be endowed with charge that produces long-range field as in the ordinary electrodynamics, but without well-localized source. This strangely-behaved charge is dubbed the {\it Cheshire charge}~\cite{Alford:1990mk, Bucher:1991bc}. Though still hypothetical, the Alice string concept found interesting applications in various areas of theoretical physics. In condensed matter, the Bose-Einstein Condensates (BEC) is known to contain the global Alice vortices~\cite{Leonhardt:2000km, Ruostekoski:2003qx,Kawaguchi:2012ii}. In cosmology, the possible role of Alice string as `mirror' particle was suggested in~\cite{Blinnikov:1982eh, Ignatiev:2003sy}. More recently, axionic Alice string is thought to solve the domain walls problem~\cite{Sato:2018nqy, Chatterjee:2019rch}. It is also worth to mention that Okada and Sakatani proposed an interesting argument that defect branes can be perceived as a realization of Alice string in string theory~\cite{Okada:2014wma}.

The simplest model that contains Alice string in its spectrum is an $SO(3)$ gauge theory with a Higgs field $\Phi$ that belongs to its $5$-dimensional irreducible representation, $\Phi\rightarrow S\Phi S^{-1}$, where $S\in SO(3)$~\cite{Bucher:1991qhl, Bucher:1990fq}. With appropriate choice of ground state, the symmetry is spontaneously broken to $SO(3)\rightarrow U(1)\rtimes\mathbb{Z}_2$. The first homotopy group of the vacuum manifold is nontrivial, $\pi_1\left(\mathcal{M}\right)\cong\mathbb{Z}_2$~\cite{Postma:1997}. This $\mathbb{Z}_2$-charge is topologically stable, but continuously deformable when any two of them collide. Consequently, two Alice strings annihilate each other. Thus, in this simplest model the Alice cannot be BPS where multi-vortices can stably exist. However, the $SO(3)$ gauge is not the only model where Alice string can be embedded. The BEC vortices discussed previously is one immediate counter-example~\cite{Leonhardt:2000km, Ruostekoski:2003qx,Kawaguchi:2012ii}, where Alice string can be found in the $U(1)\times SO(3)\rightarrow O(2)$ symmetry-breaking scenario. The homotopy group is non-trivial, $\pi_1(\mathcal{M})\cong\mathbb{Z}$. This $\mathbb{Z}$-string can exist in multi-configuration, as opposed to its $\mathbb{Z}_2$-counterparts, and thus is BPS-compatible.

Recently, Chatterjee and Nitta found the BPS states of Alice strings in the $U(1)\times SO(3)$ gauge theory coupled with charged complex scalar fields in the vector representation~\cite{Chatterjee:2017jsi}. This model is a gauged version of the BEC global Alice vortices~\cite{Leonhardt:2000km}. The Bogomolny equations are achieved by the usual Bogomolny completion of the corresponding energy integral~\cite{Bogomolny:1975de}. The BPS solutions are also numerically. Not long after that, the BPS zero modes and its effective action are discussed in~\cite{Chatterjee:2017hya}. With this BPS string's discovery, the possibility of having multi Alice vortices as well as their exotic dynamics are wide open, both in field theory as well as in SUSY embedding.

In this work, we shall extend the first-order formalism~\cite{Bazeia:2017nas} to the case of non-Abelian strings. It is then applied to obtaining BPS Alice strings for several models. In Sec.~\ref{sec: formalism} we construct the first-order formalism for non-Abelian multi-field strings. We apply the formalism to the case of Alice string in the $U(1)\times SO(3)$ gauge theory coupled with charged complex scalar fields in the vector representation in Sec.~\ref{sec: ALicestrings}. Finally we summarize our findings in Sec.~\ref{sec:summ}.

\section{First-order Formalism for Non-Abelian Multi-field Strings}
\label{sec: formalism}

Consider the most general Lagrangian containing multi Higgs and gauge fields\footnote{There can be more than one types of Higgs or gauge fields.} of co-dimension-2. The gauge fields can either be: Maxwell and $(m-1)$ Yang-Mills, or pure $m$-Yang-Mills. The Lagrangian is as follows:
\begin{equation}
		\label{gen Lag}
		\mathcal{L}=\mathcal{L}(|\Phi|, A^{\mu}_j, X,Y^{(j)}).
	\end{equation}
Some explanations are in order. Here,
\begin{equation}
\Phi\equiv\left(\phi_1, \phi_2, ....\phi_n\right), \ \ A^{\mu}_j\equiv\left(A^{\mu}_1, A^{\mu}_2,...\right)\equiv\left(a^{\mu}, A^{a\mu},...\right), 
\end{equation}
with $\phi_i$ complex scalar fields, $i=1,2,..., n$, $j=1,2,..., m$, and where $a^{\mu}$, $A^{a\mu},...$ denote the Maxwell and Yang-Mills fields, respectively. We also define
\begin{equation}
\label{Lag def}
X\equiv\Tr D_\mu\Phi^\dagger D^\mu\Phi,\ \ \ Y^{(j)}\propto\Tr G_{\mu\nu (j)}G^{\mu\nu}_{~(j)},
\end{equation}
where no summation is implied on $j$, and $G^{\mu\nu}_j$ denotes the field tensor for either Maxwell $G^{\mu\nu}_{(j)}\equiv \partial^\mu a^\nu - \partial^\nu a^\mu$ or Yang-Mills $G^{\mu\nu}_{(j)}\equiv \partial^\mu A^\nu - \partial^\nu A^\mu - ig_j [A^\mu,A^\nu]$ with $g_j$ the coupling constant of the corresponding symmetry group. Details on how many fields are involved depend on the specific models considered. Here, $\mathcal{L_M}\equiv\partial\mathcal{L}/\partial\mathcal{M}$. The corresponding Euler-Lagrange equations and energy-momentum tensors are
\begin{eqnarray}
\label{EL}
D_{\mu}\left(\mathcal{L}_XD^{\mu}\Phi\right)&=&\frac{\Phi}{2|\Phi|}\mathcal{L}_{|\Phi|},\nonumber\\
\partial_{\mu}\left(\mathcal{L}_{Y^{(j)}}G^{\mu\nu}_{(j)}\right)&=&-\mathcal{L}_{A_{\nu j}},
\end{eqnarray}
 and
 \begin{equation}
 \label{tmn}
T_{\mu\nu}=\mathcal{L}_X\Tr\left(D_{\mu}\Phi^{\dag}D_{\nu}\Phi+D_{\nu}\Phi^{\dag}D_{\mu}\Phi\right)+\sum\limits^m_j\mathcal{L}_{Y^{(j)}}G_{\mu\lambda(j)}G^{\lambda}_{\ \nu(j)}-g_{\mu\nu}\mathcal{L},
 \end{equation}
respectively. 

The ansatz chosen shall depend on the corresponding symmetry group employed. However, as long as axial symmetry and staticity are employed, they can be written as
\begin{eqnarray}
\phi_i\propto f_i(r),\ \ A^{\mu}_j\propto A_j(r). 
\end{eqnarray}
In these forms,the kinetic terms will be: 
\begin{eqnarray}
X&=&-\sum\limits^n_i\left(f_i'^2+\frac{\left(A_1+\sum\limits^m_j(-1)^{i+j}A_{j+1}\right)^2}{r^2}f_i^2\right),\nonumber\\
Y^{(j)}&=&-\frac{A_j'^2}{2g_j^2r^2},
\end{eqnarray}
where $'$ denotes the derivative wrt $r$. The alternating sign $\left(-1\right)^{j+1}$ will depend on the specific model and the number of non-Abelian gauge fields $A_j$ involved. The off-diagonal components of the energy-momentum tensor will be of the form:
\begin{equation}
T_{12}=\mathcal{L}_X\sum\limits^n_i\left(f_i'^2-\frac{\left(A_1+\sum\limits^m_j(-1)^{i+j}A_{j+1}\right)^2}{r^2}f_i^2\right)\sin2\theta.
\end{equation}
The first-order formalism relies on two fundamental assumptions: {\bf axial symmetry} and {\bf energy stability under spatial rescaling}~\cite{Bazeia:2017nas}. The axial symmetry implies 
\begin{equation}
\label{axial}
\sum\limits^n_i\left(f_i'^2-\frac{\left(A_1+\sum\limits^m_j(-1)^{i+j}A_{j+1}\right)^2}{r^2}f_i^2\right)=0.
\end{equation}
For ordinary (single-field) vortex, the condition~\eqref{axial} simply reduces to 
\begin{equation}
f'^2=\frac{a^2f^2}{r^2}.
\end{equation}
For multi-field system, however, the situation becomes tricky. In this work, we assume that each Higgs field is independent, {\it i.e.,}
\begin{equation}
\label{eq:bps1}
f_i'^2=\frac{\left(A_1+\sum\limits^m_j(-1)^{i+j}A_{j+1}\right)^2}{r^2}f_i^2.
\end{equation}
Note that this may not be the most general equations derived from~\eqref{axial}. However, in the next section we show that this yields the correct BPS conditions for the known Alice strings. Eq.~\eqref{eq:bps1} is the {\bf first} of the two Bogomolny equations. With this condition, 
\begin{equation}
\label{X1}
X=-2\sum\limits^n_i f_i'^2=-2\sum\limits^n_i\frac{\left(A_1+\sum\limits^m_j(-1)^{i+j}A_{j+1}\right)^2}{r^2}f_i^2.
\end{equation}

The second fundamental assumption, {\it i.e.,} the stationary under rescaling, is essentially the {\it Derrick's theorem} for static defects~\cite{Derrick:1964ww}. The argument is as follows. For static axially-symmetric vortices, the total energy is
\begin{equation}
 E=-2\pi\int r\ dr\ \mathcal{L}\left[\Phi_i, A_j, X, Y^{(j)}\right].   
\end{equation}
Since, under rescaling $r\rightarrow\lambda r$, the kinetic terms rescale as $X\rightarrow\lambda^2X$ and $Y^{(j)}\rightarrow\lambda^4Y^{(j)}$, the energy does also rescale as
\begin{equation}
E^{(\lambda)}=-\frac{2\pi}{\lambda^2}\int r\ dr\ \mathcal{L}\left[\Phi_i, A_j, \lambda^2X, \lambda^4Y^{(j)}\right].
\end{equation}
Derrick's theorem states that the condition for stationary implies
\begin{equation}
\frac{dE^{(\lambda)}}{d\lambda}\bigg|_{\lambda=1}=0.
\end{equation}
This results in
\begin{equation}
\label{scalc}
\mathcal{L}-\mathcal{L}_{X}X-2\sum\limits^m_j\mathcal{L}_{Y^{(j)}}Y^{(j)}=0.
\end{equation}
Taking its derivative wrt to $r$, we obtain
\begin{eqnarray}
\label{tijprime}
2\sum\limits^n_i \frac{f_i'^2}{r^2}\left[r\left(r\mathcal{L}_X f_i'\right)'-\mathcal{L}_X\left(A_1+\sum\limits^m_j\left(-1\right)^{i+j}A_{j+1}\right)^2 f_i+\frac{r^2f_i\mathcal{L}_{|\Phi|}}{2\sqrt{\sum\limits^n_{\ell=1}f_{\ell}^2}}\right]\nonumber\\
+\sum\limits^m_j\frac{A_j'}{g_j^2r^2}\left[r\left(\mathcal{L}_{Y^{(j)}}\frac{A_j'}{g_j r}\right)'-2g_j\mathcal{L}_X\sum\limits^n_i\left(A_1+\left(-1\right)^{i+j}A_{j+1}\right)^2f_i^2\right]=0.
\end{eqnarray}
Here we express $|\Phi|=\sqrt{\sum\limits^n_{\ell=1}f_{\ell}^2}$. It can be shown that when setting each square bracket to zero it gives us the field equations, as derived from the Euler-Lagrange~\eqref{EL}. By inserting Eq.~\eqref{eq:bps1} into \eqref{tijprime}, the first-order equation for $A_j$ can be extracted. Generally, it will be in the form
\begin{equation}
\label{eq:bps2}
\frac{A_j'}{r}=M\left(f_i,r\right),
\end{equation}
where $M\left(f_i,r\right)$ in general is an algebraic function of $f_i$ (and $r$) but not their derivative. Eq.~\eqref{eq:bps2} constitutes the {\bf second} Bogomolny equations.

The BPS energy can be also be calculated directly. Using the condition~\eqref{scalc}, the energy density can be re-written as
\begin{eqnarray}
\label{energydensity}
\rho&=&\mathcal{L}_{X}X+2\sum\limits^m_j\mathcal{L}_{Y^{(j)}}Y^{(j)}\nonumber\\
&=&2\mathcal{L}_X\sum\limits^n_if_i'^2+\sum\limits^m_j\mathcal{L}_{Y^{(j)}}\frac{A_j'^2}{g_j^2r^2},
\end{eqnarray}
where in the last line we have made use of Eq.~\eqref{eq:bps1}. We can define an auxiliary function $W=W(f_i, A_j)$ such that
\begin{equation}
W_{f_i}\equiv\frac{\partial W}{\partial f_i}\equiv2r\mathcal{L}_Xf_i',\ \ \ W_{A_j}\equiv\frac{\partial W}{\partial A_j}\equiv\mathcal{L}_Y\frac{A_j'}{g_j^2r}.
\end{equation}
With these definitions, the energy density~\eqref{energydensity} can be writen as
\begin{eqnarray}
\rho=\frac{1}{r}\frac{dW}{dr},
\end{eqnarray}
which implies
\begin{eqnarray}
E=2\pi\int rdr\ \rho=2\pi\ W(f_i, A_j)\bigg|_{boundary}.
\end{eqnarray}

One generic feature of various Bogomolnyi alternative formalisms~\cite{Bazeia:2008tj, Bazeia:2007df, Adam:2013hza, Atmaja:2014fha, Atmaja:2015lia, Atmaja:2015umo, Bazeia:2017nas} is that, unlike the original Bogomolny's trick~\cite{Bogomolny:1975de}, they do not start from a given potential. Instead, the form of the potential is deduced such that it satisfies the Bogomolny equations and the corresponding non-trivial topology suitable for the existence of the defects. In the next section we shall apply these mechanisms to several models of Alice strings and obtaining their corresponding BPS configurations.

\section{BPS Alice Strings}
\label{sec: ALicestrings}


This Alice Lagrangian considered by Chatterjee and Nitta~\cite{Chatterjee:2017jsi} is given by
\begin{equation}
\label{lag1}
\mathcal{L}=|\Tr D_{\mu}\Phi|^2-\frac{1}{2}\Tr F_{\mu\nu}F^{\mu\nu}-\frac{1}{4}f_{\mu\nu}f^{\mu\nu}+V,
\end{equation}
where $D_{\mu}\Phi\equiv\partial_{\mu}\Phi-iea_{\mu}\Phi-ig\left[A_{\mu},\Phi\right]$, $F_{\mu\nu}\equiv\partial_{\mu}A_{\nu}-\partial_{\nu}A_{\mu}-ig\left[A_{\mu},A_{\nu}\right]$, and $f_{\mu\nu}\equiv\partial_{\mu}a_{\nu}-\partial_{\nu}a_{\mu}$. The gauge fields $a_{\mu}$ and $A_{\mu}$ refers to the Maxwell and Yang-Mills fields, respectively. Here $V=V\left(\Phi, \Phi^{\dag}\right)$, and note that we do not specify the potential beforehand. 

The appropriate static and axially-symmetric ansatze are:
\begin{eqnarray}
\label{ansatz}
\Phi=\Phi(r,\theta)&=&\begin{pmatrix}
			0 & f_1(r) e^{in\theta}\\
			f_2(r) & 0
		\end{pmatrix},\\
a_i(r,\theta)&=&-\epsilon_{ij} \frac{x_j}{er^2}\biggl(a(r)-\frac{n}{2}\biggr),\\
A_i(r,\theta)&=&-\epsilon_{ij} \frac{x_j}{gr^2}\biggl(A(r)-\frac{n}{2}\biggr) \frac{\sigma^3}{2}.
\end{eqnarray}
In our language, $f_i=\left(f_1, f_2\right)$ and $A_j=\left(a, A\right)$, thus $Y^{(j)}\equiv\left(y, Y\right)$. Notice that the ansatz for gauge fields are slightly different from~\cite{Chatterjee:2017jsi}. The following boundary conditions are imposed to ensure regularity at the core and appropriate behavior in the vacuum:
\begin{eqnarray}
\label{bound1}
f_1(0)&=&f_2'(0)=0,\ \ \ \ \ \ \ \ f_1(\infty)=f_2(\infty)=const.\equiv\xi,\nonumber\\
\nonumber\\
a(0)&=&A(0)=\frac{n}{2},\ \ \ \ \ \ \ \ a(\infty)=A(\infty)=0.
\end{eqnarray}
In these ansatze,~\eqref{X1} can be written as
\begin{eqnarray}
\label{def explicit}
X&=&-\Biggl[f_1'^2+\frac{\bigl(a+A\bigr)^2}{r^2}f_1^2 + f_2'^2+\frac{\bigl(a-A\bigr)^2}{r^2}f_2^2\Biggr],\\
y&=&-\frac{a'^2}{2e^2r^2},\\
Y&=&-\frac{A'^2}{2g^2r^2}.
	\end{eqnarray}
The Euler-Lagrange's equations~\eqref{EL} becomes
\begin{eqnarray}
\label{EL1}
\frac{1}{r}\bigl(r\mathcal{L}_Xf_1'\bigr)' - \frac{\mathcal{L}_X\bigl(a+A\bigr)^2f_1}{r^2} + \frac{f_1}{2\sqrt{f_1^2+f_2^2}} \mathcal{L}_{|\Phi|}&=&0,\nonumber\\
\frac{1}{r}\bigl(r\mathcal{L}_Xf_2'\bigr)' - \frac{\mathcal{L}_X\bigl(a-A\bigr)^2f_2}{r^2} + \frac{f_2}{2\sqrt{f_1^2+f_2^2}} \mathcal{L}_{|\Phi|}&=&0,\nonumber\\
r\left(\mathcal{L}_y\frac{a'}{er^2}\right)' - 2e\mathcal{L}_X \left[ \left(a+A\right) f_1^2 + \left(a-A\right) f_2^2 \right]&=&0,\nonumber\\
r\left(\mathcal{L}_Y \frac{A'}{r}\right)' - 2g^2\mathcal{L}_X\Bigl[\bigl(a+A\bigr)f_1^2-\bigl(a-A\bigr)f_2^2\Bigr]&=&0.
\end{eqnarray}
The condition~\eqref{tijprime} in these ansatze can also be written as
	\begin{eqnarray}
		\label{Lag scal a}
		&&2f_1' \biggl[\frac{1}{r}\bigl(r\mathcal{L}_Xf_1'\bigr)' - \frac{\mathcal{L}_X\bigl(a+A\bigr)^2f_1}{r^2} + \frac{f_1}{2\sqrt{f_1^2+f_2^2}} \mathcal{L}_{|\Phi|}\biggr]\nonumber\\
		\label{Lag scal b}
		&&+ 2f_2' \biggl[\frac{1}{r}\bigl(r\mathcal{L}_Xf_2'\bigr)' - \frac{\mathcal{L}_X\bigl(a-A\bigr)^2f_2}{r^2} + \frac{f_2}{2\sqrt{f_1^2+f_2^2}} \mathcal{L}_{|\Phi|}\biggr]\nonumber\\
		&&+ \frac{a'}{e^2r^2} \biggl[r\bigl(\mathcal{L}_y\frac{a'}{er^2}\bigr)' - 2e\mathcal{L}_X \Bigl\{ \bigl(a+A\bigr) f_1^2 + \bigl(a-A\bigr) f_2^2 \Bigr\}\biggr]\nonumber\\
		&&+ \frac{A'}{g^2r^2} \biggl[r\bigl(\mathcal{L}_y\frac{A'}{gr^2}\bigr)' - 2g\mathcal{L}_X \Bigl\{ \bigl(a+A\bigr) f_1^2 - \bigl(a-A\bigr) f_2^2 \Bigr\}\biggr] =0.
	\end{eqnarray}
Notice that each term in the square bracket is equivalent to the lhs of each Euler-Lagrange equations above, Eq.~\eqref{EL1}. For completeness, we also show the spatial components of~\eqref{tmn} as:
\begin{eqnarray}
\label{EMT1}
T_{11}&=&\mathcal{L}_y \frac{a'^2}{e^2r^2} + \mathcal{L}_Y \frac{A'^2}{g^2r^2} + 2\mathcal{L}_X \Bigl\{\bigl(f_1'^2+f_2'^2\bigr)\cos^2\theta \nonumber\\
&&+\Bigl[\bigl(a+A\bigr)^2f_1^2+\bigl(a-A\bigr)^2f_2^2\Bigr]\sin^2\theta\Bigr\} + \mathcal{L},\nonumber\\
T_{12}&=&\mathcal{L}_X \Biggl[f_1'^2-\frac{\bigl(a+A\bigr)^2f_1^2}{r^2}+f_2'^2-\frac{\bigl(a-A\bigr)^2f_2^2}{r^2}\Biggr]\sin2\theta,\nonumber\\
T_{22} &=& \mathcal{L}_y \frac{a'^2}{e^2r^2} + \mathcal{L}_Y \frac{A'^2}{g^2r^2} + 2\mathcal{L}_X \Bigl\{\bigl(f_1'^2+f_2'^2\bigr)\sin^2\theta \nonumber\\
&&+\Bigr[\bigl(a+A\bigr)^2f_1^2+\bigl(a-A\bigr)^2f_2^2\Bigr]\cos^2\theta\Bigl\} + \mathcal{L}.
\end{eqnarray}

The first Bogomolny equation, \eqref{eq:bps1}, becomes
\begin{eqnarray}
\label{syma}
f_1'^2&=&\frac{\bigl(a+A\bigr)^2f_1^2}{r^2},\\ 
\label{symb}
f_2'^2&=&\frac{\bigl(a-A\bigr)^2f_2^2}{r^2}.
\end{eqnarray}
Deriving Eqs.~\eqref{syma}-\eqref{symb} wrt to r, and then inserting them into Eq.~\eqref{EL1} results in
\begin{eqnarray}
\label{exp a}
&&\frac{\bigl(a+A\bigr)f_1}{r} \Biggl[\mathcal{L}_{XX}X' + \mathcal{L}_{Xy}y' + \mathcal{L}_{XY}Y' + \mathcal{L}_{X\Phi} \frac{\bigl(a+A\bigr)f_1^2/r + \bigl(a-A\bigr)f_2^2/r}{\sqrt{f_1^2+f_2^2}}\Biggr]\nonumber\\
&&+\frac{a'+A'}{r}f_1\mathcal{L}_X + \frac{f_1}{2\sqrt{f_1^2+f_2^2}}\mathcal{L}_{|\Phi|}=0.
\end{eqnarray}
\begin{eqnarray}
\label{exp b}
&&\frac{\bigl(a-A\bigr)f_1}{r} \Biggl[\mathcal{L}_{XX}X' + \mathcal{L}_{Xy}y' + \mathcal{L}_{XY}Y' + \mathcal{L}_{X\Phi} \frac{\bigl(a+A\bigr)f_1^2/r + \bigl(a-A\bigr)f_2^2/r}{\sqrt{f_1^2+f_2^2}}\Biggr]\nonumber\\
&&+\frac{a'-A'}{r}f_1 \mathcal{L}_X +\frac{f_2}{2\sqrt{f_1^2+f_2^2}}\mathcal{L}_{|\Phi|} = 0.
\end{eqnarray}	
Judging from the form of the Lagrangian~\eqref{lag1}, $\mathcal{L}=X+Y+y-V$, it is not difficult to show that Eqs.~\eqref{exp a}-\eqref{exp b} reduce to the following:
\begin{eqnarray}
\label{exp1}
\frac{a'+A'}{r}f_1 + \frac{1}{2}\mathcal{L}_{f_1} = 0,\ \ \ \ \frac{a'-A'}{r}f_2 + \frac{1}{2}\mathcal{L}_{f_2} = 0.
\end{eqnarray}
Eqs.~\eqref{exp1} are the second Bogomolny equations in this model. With these conditions, \eqref{scalc} becomes
	\begin{equation}
		\label{scal1}
		-y-Y-V = 0.
	\end{equation}

Now here comes the tricky parts. First, we assume that the potential can be separated into two terms $V=V_1+V_2$. Second, each can be identified as follows:
	\begin{eqnarray}
		\label{V121a}
		V_1 &=& -y\ \ \ \ \ \ \rightarrow\ \ \  \frac{a'}{r} = e\sqrt{2V_1},\\
		\label{V121b}
		V_2 &=& -Y\ \ \ \ \  \rightarrow\ \ \  \frac{A'}{r} = g\sqrt{2V_2}.
	\end{eqnarray}
Combining all the previous equations, we have the constraint conditions
\begin{eqnarray}
\label{eV12a}
\Bigl(e\sqrt{2V_1} + g\sqrt{2V_2}\Bigr)f_1 - \frac{1}{2} V_{1,f_1} - \frac{1}{2} V_{2,f_1} &=& 0,\\
\label{eV12b}
\Bigl(e\sqrt{2V_1} - g\sqrt{2V_2}\Bigr)f_2 - \frac{1}{2} V_{1,f_2} - \frac{1}{2} V_{2,f_2} &=& 0.
\end{eqnarray}
Third, the constraints above can be satisfied by taking
\begin{eqnarray}
    \label{VV1}
    \frac{\partial V_1}{\partial f_1}&=&2e\sqrt{2V_1}f_1,\ \ \ \ \ \ \ \frac{\partial V_1}{\partial f_2}=2e\sqrt{2V_1}f_2,\\
\nonumber\\
    \label{VV2}
    \frac{\partial V_2}{\partial f_1}&=&2g\sqrt{2V_2}f_1,\ \ \ \ \ \ \ \frac{\partial V_2}{\partial f_2}=2g\sqrt{2V_2}f_2.
\end{eqnarray}
These configurations are chosen to guarantee the Schwarz's integrability condition. The solution for Eq.~\eqref{VV1} is
\begin{eqnarray}
\label{V1f1}
V_1(f_1, f_2)&=&\frac{e^2}{2}\left(f_1^2+f_2^2-2\xi^2\right)^2,   \end{eqnarray}
where the integration constant $2\xi^2$ can be identified as the VEV $\langle\Phi\rangle$. In terms of the $\Phi$ field it can be written as
\begin{equation}
V_1\left(\Phi\right)=\frac{e^2}{2}\left(\Tr\Phi^{\dag}\Phi-2\xi^2\right)^2. 
\end{equation}
Similarly, the solution for Eq.~\eqref{VV2} is
\begin{equation}
\label{V2f2}
V_2(f_1, f_2)=\frac{g^2}{2}\left(f_1^2-f_2^2+C\right)^2,    
\end{equation}
or
\begin{equation}
V_2(\Phi)=\frac{g^2}{4}\Tr\left([\Phi^{\dag},\Phi]+C\right)^2,   
\end{equation}
where $C$ is another integration constant. this time, without loss of generality we can set $C=0$. Thus, we will have the following result for the general potential that can generate BPS equations.	
	\begin{eqnarray}
		\label{Vr1}
		V(\Phi) &=& V_1(\Phi)+V_2(\Phi),\\
		&=& \frac{e^2}{2} \Bigl(\Tr \Phi^\dagger\Phi -2\xi^2\Bigl)^2 + \frac{g^2}{4} \Tr\Bigl([\Phi^\dagger,\Phi]\Bigr)^2.
	\end{eqnarray}
Note that we reproduce the same BPS potential as in~\cite{Chatterjee:2017jsi}.

\begin{figure}[h]
\centering
	\hspace{-1.5cm}
\includegraphics[scale=0.8]{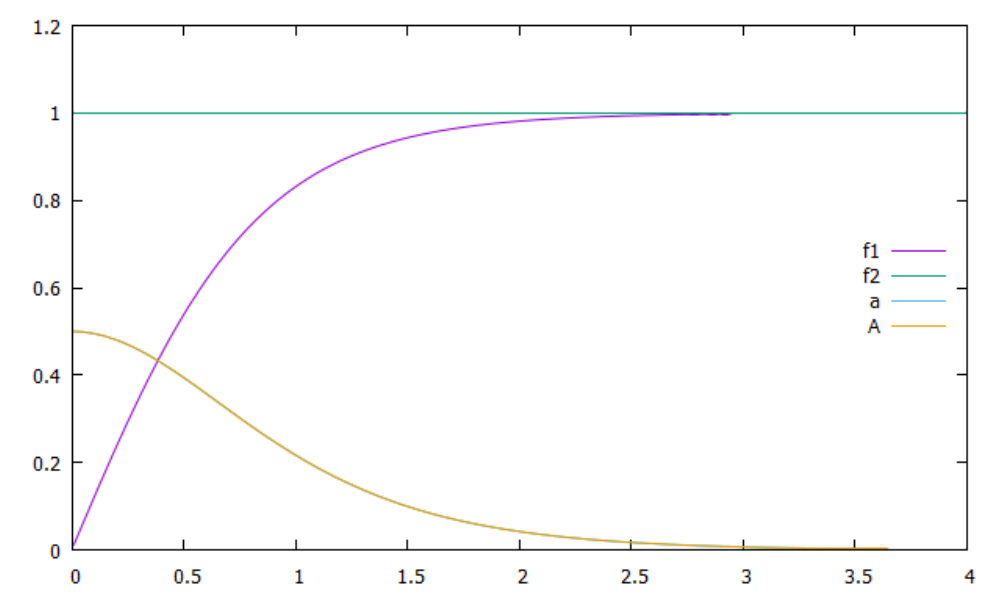}
	\caption{Profiles of scalar and gauge fields for $n=1$ and $\ell=1$.}
	\label{fig:sol1a}
\end{figure}

\begin{figure}[h]
			\hspace{-1.5cm}
\includegraphics[scale=0.8]{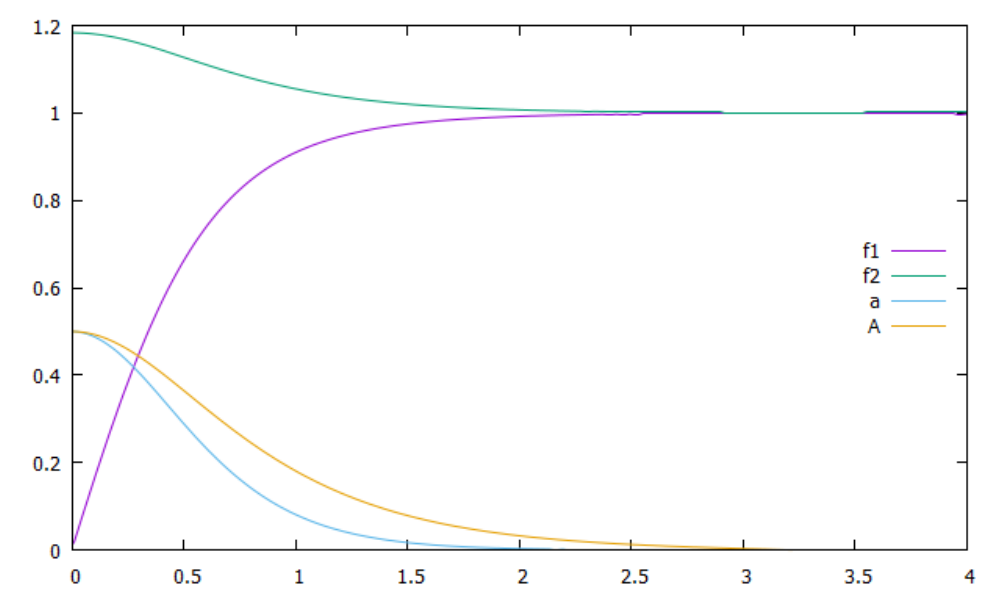}
			\caption{Profiles of scalar and gauge fields for $n=1$ and $\ell=2$.}
		\label{fig:sol2b}
\end{figure}

To summarize, by means of Eqs.~\eqref{syma},~\eqref{symb},~\eqref{V121a},~\eqref{V121b},~\eqref{V1f1}, and~\eqref{V2f2}, the BPS equations are
\begin{eqnarray}
\label{BPS1}
f_1'&=&\frac{\bigl(a+A\bigr)f_1}{r},\nonumber\\
f_2'&=&\frac{\bigl(a-A\bigr)f_2}{r},\nonumber\\
\frac{a'}{r}&=&e^2\bigl(f_1^2 + f_2^2 - 2\xi^2\bigr),\nonumber\\
\frac{A'}{r}&=&g^2\bigl(f_1^2 - f_2^2 \bigr).
	\end{eqnarray}
Imposing the boundary conditions~\eqref{bound1} we solve Eqs.~\eqref{BPS1} numerically. To get the profiles of Alice strings, we set $n = 1$. The typical results are shown in Figs.~\ref{fig:sol1a}-\ref{fig:sol2b}. As in~\cite{Chatterjee:2017jsi} for the parameter value $\ell\equiv e/g=1$, the boundary conditions force the $f_2(r)$ profile to be constant and the $a(r)$ coincides with $A(r)$, as shown in Fig.~\ref{fig:sol1a}. For other value of $\ell$ we can see, for example in Fig.~\ref{fig:sol2b}, that all profile functions interpolate distinctly between the boundaries.	

\section{Summary}
\label{sec:summ}

In this work we generalize the first-order formalism of obtaining Bogomolnyi equations to the case of non-Abelian strings. Here we neither assume any specific group nor the number of gauge fields involved. The kinetic terms are not restricted to canonical form and may take any arbitrary non-canonical function. We only require axial symmetry and staticity. As in the Abelian case developed by Bazeia \textit{et al.} \cite{Bazeia:2017nas}, this formalism relies on {\it axial symmetry} and {\it Derrick's theorem}. The former implies BPS equations for the Higgs fields, while the latter yields the equations for the gauge fields. Both must be consistent in a sense that they should satisfy the constraint equation. In this formalism, the potential is not put by hand. Rather, we start from unspecified potential and move forward by applying the algorithm to obtain the suitable BPS equations. The potential is obtained from the equation(s) of constraint. The ``tricky" part in this generalization is that it assumes that each Higgs field is independent on each other. Thus the first Bogomolny equation can be written as in Eq.~\eqref{eq:bps1}. We do not claim that this postulate is the most general we can have, though we do not have the proof to rule it out either, but certainly for the cases considered here it does work.


We apply the formalism to the $U(1)\times SO(3)$ gauge theory coupled with charged complex scalar field. Chatterjee and Nitta already derived the BPS equations for such a model~\cite{Chatterjee:2017jsi} using the usual Bogomolny's trick~\cite{Bogomolny:1975de}. Here we derive the same BPS equations using the first-order formalism, and obtain the same potential up to some constant. 
Theoretically, the scope of this work's formalism is somehow limited, since not many non-Abelian strings are topologically nontrivial. Alice strings are one of the few non-Abelian vortices with genuine topological stability, thus our choice of model studied in this work. Nevertheless, the generalized first-order formalism we develop can be applied to any arbitrary static non-Abelian strings. 
	
	
\acknowledgments
We thank Ardian Atmaja and Ilham Prasetyo for the fruitful discussion. HSR is funded by the Hibah Riset FMIPA UI No.~PKS-026/UN2.F3.D/PPM.00.02/2023.


\begin{thebibliography}{999}

\bibitem{Chatterjee:2017jsi}
C.~Chatterjee and M.~Nitta,
``BPS Alice strings,''
JHEP \textbf{09} (2017) 046
[arXiv:1703.08971 [hep-th]].

\bibitem{Bogomolny:1975de}
E.~B.~Bogomolny,
``Stability of Classical Solutions,''
Sov. J. Nucl. Phys. \textbf{24} (1976) 449, Yad.Fiz. \textbf{24} (1976) 861-870.

\bibitem{Prasad:1975kr}
M.~K.~Prasad and C.~M.~Sommerfield,
``An Exact Classical Solution for the 't Hooft Monopole and the Julia-Zee Dyon,''
Phys. Rev. Lett. \textbf{35} (1975) 760-762.


\bibitem{Bazeia:2008tj}
D.~Bazeia, L.~Losano and R.~Menezes,
``First-order framework and generalized global defect solutions,''
Phys. Lett. B \textbf{668} (2008), 246-252
[arXiv:0807.0213 [hep-th]].

\bibitem{Bazeia:2007df}
D.~Bazeia, L.~Losano, R.~Menezes and J.~C.~R.~E.~Oliveira,
``Generalized Global Defect Solutions,''
Eur. Phys. J. C \textbf{51} (2007), 953-962
[arXiv:hep-th/0702052 [hep-th]].

\bibitem{Adam:2013hza}
C.~Adam, L.~A.~Ferreira, E.~da Hora, A.~Wereszczynski and W.~J.~Zakrzewski,
``Some aspects of self-duality and generalised BPS theories,''
JHEP \textbf{08} (2013), 062
[arXiv:1305.7239 [hep-th]].

\bibitem{Atmaja:2014fha}
A.N.~Atmaja and H.S.~Ramadhan,
``Bogomol’nyi equations of classical solutions,''
Phys. Rev. D \textbf{90} (2014) no.10, 105009
[arXiv:1406.6180 [hep-th]].

\bibitem{Atmaja:2015lia}
A.N.~Atmaja, H.S.~Ramadhan and E.~da Hora,
``A Detailed Study of Bogomol'nyi Equations in Two-Dimensional Generalized Maxwell-Higgs Model Using On-Shell Method,''
JHEP \textbf{102} (2016) 117
[arXiv:1505.01241 [hep-th]].

\bibitem{Atmaja:2015umo}
A.~N.~Atmaja,
``A Method for BPS Equations of Vortices,''
Phys. Lett. B \textbf{768} (2017), 351-358
[arXiv:1511.01620 [hep-th]].
  
\bibitem{Bazeia:2017nas}
D.~Bazeia, L.~Losano, M.A.~Marques, R.~Menezes and I.~Zafalan,
``First Order Formalism for Generalized Vortices,''
Nucl. Phys. B \textbf{934} (2018) ('212-239', '212')
[arXiv:1708.07754 [hep-th]].

\bibitem{Sokalski:2001wk}
K.~Sokalski, T.~Wietecha and Z.~Lisowski,
``A concept of strong necessary condition in nonlinear field theory,''
Acta Phys. Polon. B \textbf{32} (2001), 2771-2792

\bibitem{Nielsen:1973cs}
H.~B.~Nielsen and P.~Olesen,
``Vortex Line Models for Dual Strings,''
Nucl. Phys. B \textbf{61} (1973), 45-61

\bibitem{Kneipp:2001tp}
M.~A.~C.~Kneipp and P.~Brockill,
``BPS string solutions in nonAbelian Yang-Mills theories,''
Phys. Rev. D \textbf{64} (2001), 125012
[arXiv:hep-th/0104171 [hep-th]].

\bibitem{Kneipp:2002yv}
M.~A.~C.~Kneipp,
``BPS Z(k) strings, string tensions and confinement in nonAbelian theories,''
PoS \textbf{unesp2002} (2002), 020
[arXiv:hep-th/0211146 [hep-th]].

\bibitem{Kneipp:2002gk}
M.~A.~C.~Kneipp,
``Z(k) string fluxes and monopole confinement in nonAbelian theories,''
Phys. Rev. D \textbf{68} (2003), 045009
[arXiv:hep-th/0211049 [hep-th]].

\bibitem{Kneipp:2008dc}
M.~A.~C.~Kneipp,
``Hitchin's equations and integrability of BPS Z(N) strings in Yang-Mills theories,''
JHEP \textbf{11} (2008), 049
[arXiv:0801.0720 [hep-th]].

\bibitem{Vachaspati:1991dz}
T.~Vachaspati and A.~Achucarro,
``Semilocal cosmic strings,''
Phys. Rev. D \textbf{44} (1991), 3067-3071

\bibitem{Vachaspati:1992jk}
T.~Vachaspati,
``Electroweak strings,''
Nucl. Phys. B \textbf{397} (1993), 648-671

\bibitem{Vilenkin:2000jqa}
A.~Vilenkin and E.~P.~S.~Shellard,
``Cosmic Strings and Other Topological Defects,''
Cambridge University Press, 2000,
ISBN 978-0-521-65476-0
		
    \bibitem{Schwarz:1982ec}
    A.~S.~Schwarz,
    ``Field theories with no local conservation of the electric charge,''
    Nucl. Phys. B \textbf{208} (1982), 141-158.

\bibitem{Benson:2004ue}
K.~M.~Benson and T.~Imbo,
``Topologically Alice strings and monopoles,''
Phys. Rev. D \textbf{70} (2004), 025005
[arXiv:hep-th/0407001 [hep-th]].

\bibitem{Bucher:1991qhl}
M.~Bucher, H.~K.~Lo and J.~Preskill,
``Topological approach to Alice electrodynamics,''
Nucl. Phys. B \textbf{386} (1992), 3-26
[arXiv:hep-th/9112039 [hep-th]].

\bibitem{Alford:1990mk}
M.~G.~Alford, K.~Benson, S.~R.~Coleman, J.~March-Russell and F.~Wilczek,
``The Interactions and Excitations of Nonabelian Vortices,''
Phys. Rev. Lett. \textbf{64} (1990), 1632
[erratum: Phys. Rev. Lett. \textbf{65} (1990), 668].

\bibitem{Bucher:1991bc}
M.~Bucher, K.~M.~Lee and J.~Preskill,
``On detecting discrete Cheshire charge,''
Nucl. Phys. B \textbf{386} (1992), 27-42
[arXiv:hep-th/9112040 [hep-th]].

\bibitem{Leonhardt:2000km}
U.~Leonhardt and G.~E.~Volovik,
``How to create Alice string (half quantum vortex) in a vector Bose-Einstein condensate,''
Pisma Zh. Eksp. Teor. Fiz. \textbf{72} (2000), 66-70
[arXiv:cond-mat/0003428 [cond-mat]].

\bibitem{Ruostekoski:2003qx}
J.~Ruostekoski and J.~R.~Anglin,
``Monopole core instability and Alice rings in spinor Bose-Einstein condensates,''
Phys. Rev. Lett. \textbf{91} (2003), 190402
[erratum: Phys. Rev. Lett. \textbf{97} (2006), 069902]
[arXiv:cond-mat/0307651 [cond-mat]].

\bibitem{Kawaguchi:2012ii}
Y.~Kawaguchi and M.~Ueda,
``Spinor Bose-Einstein condensates,''
Phys. Rept. \textbf{520} (2012), 253-381.

\bibitem{Blinnikov:1982eh}
S.~I.~Blinnikov and M.~Y.~Khlopov,
``ON POSSIBLE EFFECTS OF 'MIRROR' PARTICLES,''
Sov. J. Nucl. Phys. \textbf{36} (1982), 472
ITEP-11-1982.

\bibitem{Ignatiev:2003sy}
A.~Y.~Ignatiev and R.~R.~Volkas,
``Mirror matter,''
[arXiv:hep-ph/0306120 [hep-ph]].

\bibitem{Sato:2018nqy}
R.~Sato, F.~Takahashi and M.~Yamada,
``Unified Origin of Axion and Monopole Dark Matter, and Solution to the Domain-wall Problem,''
Phys. Rev. D \textbf{98} (2018) 4, 043535
[arXiv:1805.10533 [hep-th]].

\bibitem{Chatterjee:2019rch}
C.~Chatterjee, T.~Higaki and M.~Nitta,
``Note on a solution to domain wall problem with the Lazarides-Shafi mechanism in axion dark matter models,''
Phys. Rev. D \textbf{101} (2020) 7, 075026
[arXiv:1903.11753 [hep-th]].

\bibitem{Okada:2014wma}
T.~Okada and Y.~Sakatani,
``Defect branes as Alice strings,''
JHEP \textbf{03} (2015), 131
[arXiv:1411.1043 [hep-th]].

\bibitem{Bucher:1990fq}
M.~A.~Bucher,
``On the theory of non-Abelian vortices and cosmic strings,''
UMI-91-13594.

		\bibitem{Postma:1997}
		M.~M.~H.~Postma,
		``Alice Electrodyanmics,''
		Master's Thesis (1997),
		Amsterdam U.

\bibitem{Chatterjee:2017hya}
C.~Chatterjee and M.~Nitta,
``The effective action of a BPS Alice string,''
Eur. Phys. J. C \textbf{77} (2017) no.11, 809
[arXiv:1706.10212 [hep-th]].


\bibitem{Derrick:1964ww}
G.~H.~Derrick,
``Comments on nonlinear wave equations as models for elementary particles,''
J. Math. Phys. \textbf{5} (1964), 1252-1254



 






  





\end{thebibliography}
\end{document}